\documentclass[prl,a4paper,twocolumn,showpacs,preprintnumbers,amsmath,amssymb%
nobibnotes,nofootinbib,flloatfix,amstex]{revtex4}
\newcommand{\ket}[1]{|\,#1\,\rangle}        		%
\newcommand{\bra}[1]{\langle\,#1\,}         		%
\newcommand{\fxf}{F\otimes F^{\dagger}}			
\newcommand{\De}{\text{\textbf{\textsf{D}}}_{\epsilon}}	
\newcommand{\qL}{\text{\textbf{\textsf{L}}}}		%
\newcommand{\doll}{\text{\textbf{\textsf{L}}}_{\epsilon}}%
\newfont{\Bb}{msbm10}					%
\newcommand{\oT}{\hat{T}}				%
\usepackage{graphicx,graphics,wrapfig,rotating}	
\usepackage{dcolumn}				
\usepackage{bm}					

\begin{document}

\title{MODELS OF DIFFUSIVE NOISE ON THE SPHERE }
\author{M.E.Spina $^{a}$}
\email{spina@tandar.cnea.gov.ar}

\author{M. Saraceno $^{a,b}$ }
\email{saraceno@tandar.cnea.gov.ar}

 \affiliation{%
$^{a}$ Dto. de F\'\i sica, Comisi\'on Nacional de Energ\'\i a At\'omica.
Libertador 8250 (1429), Buenos Aires, Argentina.\\
$^{b}$Escuela de Ciencia y Tecnolog\'\i a, Universidad Nacional de San
Mart\'\i n. Alem 3901 (B1653HIM), Villa Ballester, Provincia de  Buenos
Aires,
Argentina.}%

\date{\today}

\begin{abstract}
We analyze Haake et al. method \cite{Haa} for coarse graining quantum maps on the sphere from the point of view of realizable physical quantum operations achieved with completely positive superoperators. We conclude that sharp truncations \`a la Haake do not fall into this class. 
\end{abstract}

\maketitle

The dissipative kicked top has been extensively studied by Haake and collaborators  \cite{Haa}.
 In their work, the Ruelle Pollicot resonances  \cite{ruel} which rule the asymptotic decay of time-dependent quantities are evaluated both classically and quantum mechanically.  The coarse graining is 
implemented by a truncation of the evolution operator ( Perron Froebenius or Husimi propagator) to a finite dimension. The procedure consists in introducing a basis of functions ordered by resolution in phase space and then performing a truncation of dimension $ N $. The eigenvalues of the $N$-dimensional matrix are evaluated and the the limit $  N \rightarrow \infty $ is finally taken. 

An alternative way of implementing dissipation in decoherent quantum maps is via a coarse-graining superoperator.  Such a model has been introduced  in \cite{123} for chaotic maps on the torus.  In this 
approach the Ruelle resonances are determined as the eigenvalues of a superoperator which is the composition of   
a unitary and a diffusive part, the latter being represented by a Kraus sum \cite{krau}. 

In this letter the superoperator formalism will be extended to maps on the sphere, more particularly to the 
kicked top, by appropiately modelling the diffusive step on the sphere. Our purpose is to study the relationship between this  approach and the {\it sharp truncation formalism} of \cite{Haa}. That is, we will investigate whether a truncation procedure  \` a la Haake can be implemented via a  diffusive superoperator, and which specific form such a procedure would imply for it. Reciprocally, it will be interesting to see how a given modelisation of  the Kraus superoperator leads to a different criterium for the basis truncation in the {\it sharp truncation formalism}.\\

The dynamics of the kicked top is given by an area preserving map, consisting of rotation and torsion operators, acting on a vector $ \hat{J} = j \ (\sin \theta  \ \cos \phi, \sin \theta \ \sin \phi ,\cos \theta) $ of fixed length $ j $, $ j $ playing the role of the inverse Planck's constant. The corresponding phase space is the sphere, and $ \cos \theta $ and $ \phi $ the canonical variables.

\noindent Its quantal version is specified by a Floquet operator: 

\begin{equation}
F = R_z (\tau cos \theta) \ R_z (\beta_z) \ R_y (\beta_y)
\end{equation}

\noindent where $ R_i (\beta) $ is a rotation around axis $ i $ of an angle $ \beta $.
The Hilbert space of the wave functions is spanned by the $ ( 2 j + 1 ) $ eigenvectors of $ \hat{J_z}  $, $ \ket {jm} $ . 
Since we are dealing with  open systems states will be represented by a density operator $ \hat{\rho} $. 
 In \cite{Haa} this density operator $ \hat{\rho} $ is represented by the corresponding Husimi function 
$ H_{\rho} ( \theta, \phi) = \bra{j \theta \phi} | \hat{\rho}  \ket{j \theta \phi} $ , that is, its diagonal matrix element in the basis of coherent states , which in turn is expanded in terms of spherical harmonics:

\begin{equation}
\label{trun}
H_{\rho} ( \theta, \phi) = \sum_{L=0}^{2j}  \sum_{M=-L}^{M=L}  H_{\rho} (L,M) Y_{L M} (\theta, \phi)
\end{equation}

\noindent The function  $ H_{\rho} (L,M) $ can be thought as the representation of $ \hat{\rho} $ in an orthogonal basis of operators $ \oT_{LM} $ given by $ \oT_{LM} =\int d \cos \theta  \ d \phi \  \ket{j \theta \phi}  Y_{L}^{M} (\theta, \phi)  \bra{j \theta \phi}|  $. These are related to the usual angular momentum states through:

\begin{eqnarray}
\oT_{LM}  & = & (-)^j \sqrt{{4 \pi } \over (2 L + 1)}  C_{j j j -j}^{L 0}  \nonumber\\
& & \sum_{m_1,m_2}  (-)^{m_1} C_{j -m_1 j m_2}^{L M} 
 \ket{j m_1}    \bra{j m_2}|
\label{tlm}
\end{eqnarray}

\noindent where  $ C_{j -m j' m'}^{J M} $ are Clebsch Gordan coefficients.
The coarse graining is then introduced by fixing a resolution parameter $L_{max} < 2 j $ and truncating  the Husimi propagator $ U $ of matrix elements 

\begin{equation}
U_{LM,L'M'} = tr (\oT_{LM}^{\dagger} F  \oT_{L'M'} F^{\dagger})
\label{husi}
\end{equation}

\noindent to be diagonalized,  to a dimension $ ( L_{max}+1)^2 $. If the area of the sphere is normalized to $ 4 \pi $ then  the finest detail of the quantum distribution is the ' sub-planck' cell \cite{zur} of size  $ 4 \pi  / (2 j + 1)^2 $. A truncation of size $ L_{max} $ eliminates all structures of size finer than 
$ 4 \pi / ( L_{max}+ 1)^2 $.

We now consider the superoperator formalism. Here 
the linear action of $ F $ on the space of density matrices defines a unitary superoperator $ \qL = \fxf $, of dimension $ (2j + 1)^2 \times  (2j + 1)^2 $.
Interaction with the environment and decoherence is then introduced by composing $ \qL $ with  a diffusion superoperator $ \De $, so that the form of the full propagator is : $ \doll = \De   \circ \qL $. In analogy with the superoperator defined in \cite{123} for maps on the torus,  which was a superposition of translation operators, $ \De $ is here an incoherent superposition of  rotations $ R( \Omega) $, each rotation having a probability 
$ c_{\epsilon}  ( \Omega) $.  Its Kraus representation \cite{krau} reads :

\begin{equation}
\De = \int d \Omega  \  c_{\epsilon} ( \Omega) R(\Omega)  \otimes R^{\dagger}(\Omega).
\label{angle}
\end{equation}

\noindent and its action on the density matrix will be:

\begin{equation}
\De \rho= \int d \Omega  \  c_{\epsilon} ( \Omega) R(\Omega)  \hat{\rho} R^{\dagger}(\Omega).
\label{rho}
\end{equation}

\noindent The preservation of the trace implies that:

\begin{eqnarray}
  \int d \Omega  \  c_{\epsilon} ( \Omega) & = &1 \nonumber \\
  c_{\epsilon} ( \Omega)  & \geq  & 0  . 
\label{posi}
\end{eqnarray}

\noindent In addition,  an overall drift in any particular direction should be avoided, leading to a further constraint on the form of $ c_{\epsilon}  ( \Omega) $.  If the rotations are expressed as $ R (\omega, \Theta,\Phi) $, that is, as rotations through an angle $ \omega $ around some axis $ \hat{n} (\Theta, \Phi) $, isotropy requires  that all directions of  $\hat{n}$ should be equally probable and thus 
that the coefficient  $c_{\epsilon} ( \Omega) $ should be independent of the direction of $\hat{n}$ and only depend on $ \omega $.  
Furthermore if we take  $c_{\epsilon} ( \omega) $  to be a narrow function peaked at $ \omega = 0$, the action of $ \De $ consists in displacing any state incoherently and isotropically over a region of area $ \propto \epsilon^2 $ on the surface of the sphere.  The coarse graining parameter   $ \epsilon $ is then proportional to $ L_{max} ^{-1} $, $ L_{max} $ being the resolution parameter defined above.

An alternative modelisation of the diffusion superoperator can be given in terms of the operators   $ \oT_{LM}  $ introduced in eq.(\ref{tlm}) as:

\begin{equation}
\De ^{\prime} = \sum_{L M}  \  c_{\epsilon}^{L}  \oT_{LM}   \otimes \oT_{LM}^{\dagger}.
\label{alter}
\end{equation}

\begin{figure}[h]
\begin{center}
\hspace{-0.5cm}
\includegraphics*[width=7cm,angle=-90]{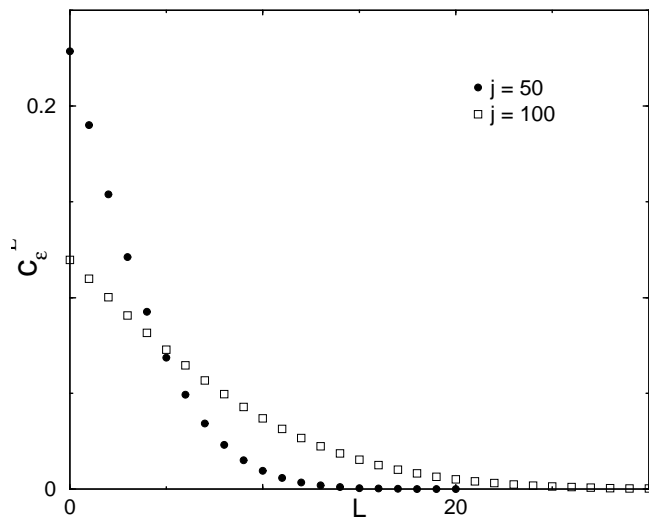}
\vspace{-0.25cm}
\caption{ Normalized coefficients $  c_{\epsilon}^{L} $ corresponding to  $ c_{\epsilon} ( \omega) $ given by eq.(\ref{cegau}) for $ \epsilon = 0.1 $ and two values of j.
\label{pl1}}
\end{center}
\end{figure}

\noindent In this case the preservation of the trace implies that:

\begin{equation}
\sum_{L=0}^{2 j} c_{\epsilon}^{L}  (C_{j j j -j}^{L 0})^2= {(2 j +1) \over 4 \pi}
\end{equation}

\noindent while the isotropy requirement is satisfied by imposing that the coefficients $ c_{\epsilon}^L $ should not depend on the angular momentum projection $ M $. 
Under this condition one can show that $ \De ^{\prime} $ coincides with $ \De $ if the $ c_{\epsilon}^{L} $ are related to the  $c_{\epsilon} ( \omega) $ through:

\begin{eqnarray}
c_{\epsilon}^{L} &=& 2 (2 L +1) {(-)^L \over (C_{j j j -j}^{L 0})^2} \sum_K (-)^K
\left\{
\begin{array} {ccc}
   j & j & K \\
   j & j & L
\end{array}
\right\} \nonumber\\
& & \int d \omega   \sin^2  {\omega  \over 2} c_{\epsilon} ( \omega)  \chi^K (\omega)
\label{relat}
\end{eqnarray}

\noindent where $  \left\{
\begin{array} {ccc}
   j & j & K \\
   j & j & L
\end{array}
\right\} $ indicates the $ 6j $ symbol and  $ \chi^K(\omega)$ is the character of the $ SU(2) $ irreducible representation of rank $K$.
It is then clear that the diffusion superoperator can be specified indistinctly either by modelling the coefficients  $ c_{\epsilon} ( \omega) $ or the $ c_{\epsilon}^{L} $.  For example, taking $c_{\epsilon} ( \omega) $  to be a  gaussian - like function of width $ \epsilon $ peaked at $ \omega = 0$ corresponds to a  $ c_{\epsilon}^{L} $ peaked at $ L = 0 $ and having a width proportional to $ j  \epsilon $, as shown in Fig. (\ref{pl1}). 
\\

In order to find the link between this model and the {\it sharp truncation formalism}, the spectral properties of $ \De $ have to be analyzed. These are not as simple as in the case of maps on the torus, where the eigenfunctions of the translation operators were the translation operators themselves.  Here the eigenstates of $ \De $ depend on the particular form of the coefficients  $c_{\epsilon} ( \Omega) $. Nevertheless, it can be shown that, provided  $ \De $ is isotropic, we have:

\begin{eqnarray}
 (\oT_{L' M'}, \De \oT_{LM}) & = & \delta_{L L'}  \delta_{M M'} |  \oT_{LM} |^2  {1 \over (2 L +1)}     \nonumber\\
& & \ \int d \omega  \sin^2  {\omega  \over 2} c_{\epsilon} ( \omega)  \chi^L (\omega).
\end{eqnarray}

\noindent That is, the operators  $\oT_{LM} $ are eigenstates of $ \De $, their corresponding eigenvalues being:

\begin{equation}
e^L_{\epsilon} = {8 \pi \over (2 L +1)}   \int d \omega  \sin^2  {\omega  \over 2} c_{\epsilon} ( \omega)  \chi^L (\omega)
\label{cele}
\end{equation}

\noindent indepent of $ M $.
The fact that $\De$ is diagonal in the  $ \{\oT_{LM} \} $ - basis,  which is precisely the one chosen in \cite{Haa} to achieve the truncation, enables us to  write the matrix elements of the full propagator $ \doll $ as:

\begin{equation}
(\doll)_{LM,L'M'} = e^L_{\epsilon}  tr (\oT_{LM}^{\dagger} F  \oT_{L'M'} F^{\dagger})
\end{equation}

\noindent and compare this expression with the one corresponding to the matrix elements of the truncated Husimi propagator given in eq.(\ref{husi}).  It becomes then clear that the truncation  of  \cite{Haa} can be reproduced in our scheme by requiring : 

\begin{equation}
e^L_{\epsilon} =  \Theta (L_{max}-L).
\label{teta}
\end{equation}

\noindent In this way, the action of the diffusion superoperator $ \De $ will be to suppress terms with $ L > L_{max} $ leaving the others unchanged, or, in other words, to reduce the $ ( 2 j + 1)^2 $ -dimensional  $  \qL $ to the  $ (L_{max} + 1)^2 $ -dimensional $ \doll $. 

\noindent Inverting  eq.(\ref{cele}), together with eq.(\ref{teta}) we get the form of the $ c_{\epsilon} ( \omega) $ leading to such a spectrum:

\begin{equation}
c_{\epsilon} ( \omega)= {{[ (2 L_{max} +3) 
\chi^{L_{max}} (\omega) - (2 L_{max} +1) \chi^{L_{max}+1} (\omega)]} \over {32 \pi^2 sin^2 {\omega \over 2}}}.
\end{equation}

\begin{figure}[h]
\begin{center}
\hspace{-0.5cm}
\includegraphics*[width=7cm,angle=-90]{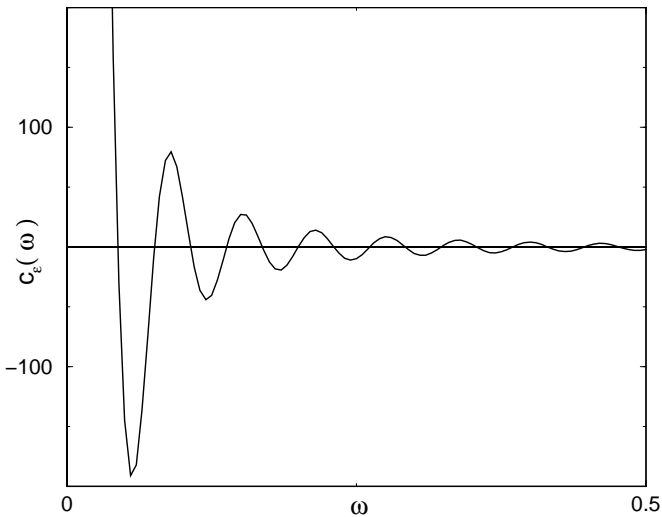}
\vspace{-0.25cm}
\caption{  Coefficient  $ c_{\epsilon} ( \omega) $ corresponding to the sharp truncation. $ L_{max} = 100$.
\label{pl2}}
\end{center}
\end{figure}

\noindent In Fig. (\ref{pl2}), $ c_{\epsilon} ( \omega) $ for $ L_{max} = 100 $ is plotted.

\noindent We see that $ c_{\epsilon} ( \omega) $ is a non positive strongly oscillating function of $ \omega $. Therefore, we can conclude that a sharp truncation cannot be achieved within  the superoperator formalism since it implies coefficients in the Kraus representation which do not fulfill the positivity requirement of eq.(\ref{posi}). \\

\noindent We could also follow the inverse procedure,  that is, start from a coefficient $ c_{\epsilon} ( \omega) $ which satisfies the superoperator requirements and use eq.(\ref{cele}) to compute the corresponding eigenvalues of  $ \De $. For example, taking $ c_{\epsilon} ( \omega) $ to be a smooth positive function of $ \omega $ :

\begin{equation}
 c_{\epsilon} ( \omega) = {1 \over 8 \pi sin^2 {\omega \over 2}} \sqrt{2 \over \pi} {1 \over {\epsilon}}  \exp [{-\sin^2 \omega \over { 2 \epsilon^2}}]
\label{cegau}
\end{equation}

\noindent we get:

\begin{equation}
e^L_{\epsilon} = {2 \over {(2L +1)}} \left( {1 \over 2} + \sum_{k=0}^{L}  \exp[{-{1 \over 2} k^2 \epsilon^2}] \right)
\end{equation}

\noindent In Fig.(\ref{pl3}) these eigenvalues are plotted for $ \epsilon = 0.1 $. We observe that the action of $ \De $ is a suppression of the high L components,  but now following a smooth function.  The width of this function is proportional to $ 1 / \epsilon  $ and its rate of decay is small, so that there are still non negligible contributions for $ L_{max} \leq L \leq 2j $. \\

We summarize our results: there are two ways to write down a Kraus
form for a diffusive superoperator on the sphere, given by eq.({\ref{angle}) and eq.({\ref{alter}).
They both represent the same completely positive superoperator provided the
coefficients are related by eq.({\ref{relat}). This superoperator is diagonal in the $T_{LM} $ 
operator basis and the eigenvalues can be tailored to truncate and
coarse-grain unitary maps on the sphere. However sharp truncations result
in linear actions that are not interpretable as {\it physical quantum
operations} derived from unitary interactions with an environment - i.e. are
not expressed as completely positive superoperators. 
This is a diffraction effect caused by the sharp edge in $L$-space.
It casts a
different light on Haake's group results but of course has no relevance to
their calculation of Ruelle resonances, which only depend on the suppression
of high frequencies in the quantum distribution.

\begin{figure}[h]
\begin{center}
\hspace{-0.5cm}
\includegraphics*[width=7cm,angle=-90]{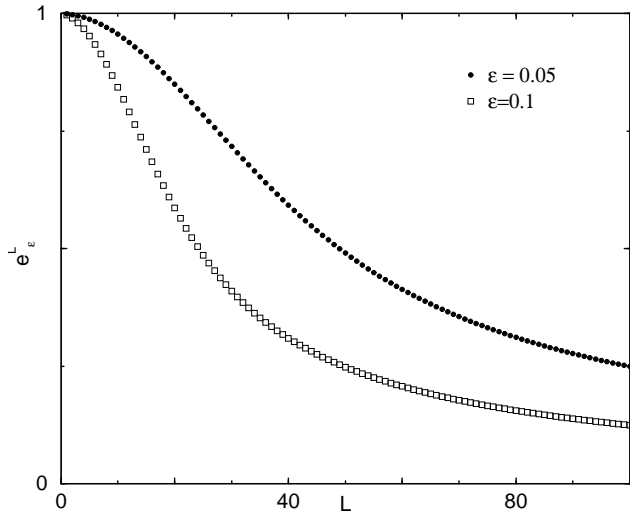}
\vspace{-0.25cm}
\caption{Eigenvalues of $ \De $ corresponding to a coefficient $ c_{\epsilon} ( \omega) $ given by eq.(\ref{cegau}) for $ \epsilon = 0.05 $ and $ \epsilon = 0.1 $.
\label{pl3}}
\end{center}
\end{figure}


\begin{thebibliography}{99}

\bibitem{Haa}
Weber J., Haake F. and Seba P.  2000 {\it Phys. Rev. Lett. } \textbf{85} 3620 ;
Weber J., Haake F., Braun P., Manderfeld C. and Seba P. 2001 {\it J.Phys} A \textbf{34} 7195 ;
 Manderfeld C., Weber J. and  Haake F.  2001 {\it J.Phys} A \textbf{34} 9893 

\bibitem{ruel}
 Ruelle D  1986 {\it Phys. Rev. Lett. } \textbf{56} 405 ;
 Ruelle D.  {\it J. Stat. Phys} \textbf{44} 281 

\bibitem{123}
Bianucci P., Paz J.P. and Saraceno M.  2002 {\it Phys. Rev.} E  \textbf{65} 046226  ;
Garc\'{\i}a-Mata I., Saraceno M. and Spina M.E.  2003 {\it Phys.Rev.Lett.} \textbf{91} 064101 ;
Garc\'{\i}a-Mata  I., Saraceno M.  2004 {\it Phys.Rev.} E \textbf{69} 056211

\bibitem{zur}
Zurek W.H.  2001 {\it Nature} \textbf{412} 712

\bibitem{krau}
Kraus K. 1983 {\it States, Effects and Operations} Springer Verlag, Berlin.
\end{thebibliography}
\end{document}